# Understanding the Mechanics of Some Localized Protocols by Theory of Complex Networks

Chiranjib Patra [a], Samiran Chattopadhyay [b], Matangini Chattopadhyay [c], Parama Bhaumik [b]

[a] *Department of Information Technology , Calcutta Institute of Engineering and Management , Tollygunge, Kolkata - 40, India*

[b] *Department of Information Technology Jadavpur University,Kolkata-32, India*

[c] *School of Education Technology Jadavpur University,Kolkata-32, India*

**Abstract**

In the study of ad hoc sensor networks, clustering plays an important role in energy conservation therefore analyzing the mechanics of such topology can be helpful to make logistic decisions .Using the theory of complex network the topological model is extended, where we account for the probability of preferential attachment and anti preferential attachment policy of sensor nodes to analyze the formation of clusters and calculate the probability of clustering. The theoretical analysis is conducted to determine nature of topology and quantify some of the observed facts during the execution of topology control protocols. The quantification of the observed facts leads to the alternative understanding of the energy efficiency of the routing protocols.

## 1. Introduction

In selecting cluster-heads, when a probabilistic method is used then each node elects itself as a cluster-head with the same probability in large-scale and homogenous WSNs because it enables all nodes to independently decide their roles while keeping the signalling overhead low. This method ensures rapid clustering while achieving favourable properties such as stable number of clusters and rotation of the cluster-heads. To distribute evenly the energy load among the nodes, the re-selection of the cluster-heads is done at a regular interval .By this practice we can extend the lifetime and minimize the energy consumption.

In carrying through the analysis of the system, the nodes are allowed to join the network through preferential attachment and leave the network or not join the network through non-preferential attachment [8] .Further the nodes distinguish themselves as cluster head nodes and normal nodes consistent with the heuristic definitions implicit in existing random graph approaches.

The goal of the paper is to quantify some of the observed features of localized Topology Control protocols (simple tree, CDS-Rule K, A3 ) used in sensor networks with the theory of complex networks.

The paper is organized as follows section 2 deals with the related work, section 3 describes the new local world model ,section 4 gives the organised analysis of the dynamical equation, section 5 extends the usage of dynamical equation for the analysis of localised TC protocols, section 6 concludes the paper.

## 2. Related Work

The B-A scale-free network model defines the basic structure that is responsible for the power-law degree distribution, it is still a approximate model with several limitations. In [5], Li Xiang and Chen Guanrong proposed a Li-Chen model to modify a limitation in the B-A model. They suggested that there should be a local world of each node in various real-world complex networks. Moreover, the preferential attachment mechanism of a scale-free net-work does not work in a global network but does work in the local world of each node. The Li-Chen model represents a transition between a power-law and an exponential distribution.

As for the formation of widely existent scale-free phenomenon in real networks, preferential connection is commonly been considered as a key factor. Taking preferential connection as a foothold, many variations of the scale-free network model is been proposed during the recent past years. For instance, the comprehensive multi-local-world model [12] which requires no global information while behaves a novel topological feature as not being complete random or complete scale-free but being somewhere between them. The local preferential attachment model [13] that is based on the common senses that people can own information easily from its local community than from outside environment. The physical position neighbourhoods model [14] which mimics the actual communication network and appears a degree distribution interpolate the power-law scaling and exponential scaling. The Poisson growth model [15] which sets up the number of edges added at each step as a random variable corresponds with Poisson distribution and can generate many kind networks by controlling the random number.

Chen et al. [6] studied a new evolving mechanism for deducing the fault-tolerant communication topology among the cluster heads with complex network theory. Based on the B-A model's growth and preferential attachment element, they not only used a local-world strategy for the network when a new node was added to its local-world but also selected a fixed number of cluster heads in the local world, for the purpose of obtaining a good performance in terms of random error tolerance.

Luo et al [8] studied the Theoretical analysis and numerical simulation conducted to explore the topology characteristics and network performance with different node energy distribution. The result shows that, when nodes energy is more heterogeneous, the network is better clustered and enjoys higher performance in terms of the network efficiency and the average path length for transmitting data.

In this paper, an evolving model for WSN's based on the B-A model and the Li-Chen model [7] is proposed. The primary goal of the paper is to find the effect of interaction of different types of sensor nodes (normal, cluster head ) in WSN's , the proportion of a normal node in a cluster node's neighbour nodes is increased in this model.

## 3. The new local world network model

In order to analyse the clustering of the sensor networks, Li-Chen model [5] is used to form the generalised local world model which forms the basis of clustering, where each node has only one local connection information; nodes connect only in their local world based on local connection information. The following are parameters required to explain the dynamics

1. Staring from a small number $\mathbf{m_0}$ of nodes at each time step **t**.

2. When choosing the nodes to which the new node connects, assume that the probability (**$k_i$**) that a new node is connected to node **i** depends on the degree **$k_i$** of node **i**, in such a way that

$$\prod(k_i) = \frac{k_i}{\sum_j k_j}$$

3. Select M nodes randomly from the existing network referred to as local world of the new coming node.

4. Add a new node with m edges , linking to m nodes in the local world determined in (3) using preferential attachment with probability **$\prod_{local}(ki)$** defined at every time step t by

$$\prod_{local}(ki) = \prod'(i \in_{local\text{-}world}) * (ki) / \left( \sum_j Local.k_j \right)$$

Where **$\prod'(i \in_{local\text{-}world}) = M/(m_0+t)$**

then include a node with **m** edges connected to the network. After **t** time steps , there will be a network with **$N=t+m_0$** nodes and **m*t** edges

This model contains two types of nodes: normal nodes and cluster nodes. There is only one cluster node attached to a normal node; in other words, the normal node has only one edge, which means that the normal node cannot relay data from other nodes. A cluster node can integrate and transmit data from other nodes. Both of the two types of nodes can connect to a cluster node, and the number of edges is limited in every cluster node because of its energy efficiency. As every new cluster node joins the network, it is randomly assigned an initial energy **$E_i$** from [0.5,1]. The limited number of edges in every cluster node is represented by **$k_{max\,i}$** , which is based on the initial energy of the cluster nodes **$E_i$** , and [0.5,1] is chosen as the interval [**$E_{min}$, $E_{max}$**] . where **$k_{max}$** [9] is given as follows:

$$K_{max\,i} = k_{max} * E_i / E_{max}$$

**$K_{max\,i}$** reflects the ability of having the maximum number of edges for cluster node ***i***.

Then, the growth model is described as follows: starting with a small number of nodes, (all of them are cluster nodes), they randomly link each other. This results into an initial network.

1. **Growth**: At every time step, a new cluster node or a normal node with one edge enters into the existing network with a probability ***p*** or **1-p** respectively .New node is a cluster node, then gives it a random energy value **$E_i$**. A small number of cluster nodes would cause many sensor nodes to link to them, which results in working frequently and consuming energy fast; however, a large number of cluster nodes would waste resources and decrease the energy efficiency. Thus, we assume that ***$0<p<0.5$***.
2. **Preferential Attachment***:* The new incoming node links to an old cluster node that is selected randomly from the pre-existing network. Nodes in WSNs have the constraint of energy and connectivity and only communicate data with the cluster nodes in their local area. First, **M** cluster nodes are selected randomly from the network as the new incoming node's local world; then, one of the cluster nodes is chosen to link with the new node according to the probability $\prod_{local}(k_i)$.

1. If the new incoming node is a cluster node, then the probability is set as follows:

$$\prod_{k_i} = \left(1-\frac{k_i}{k_{\max i}}\right)\frac{k_i}{\sum_{j\in local} k_j} \text{ ---------------------(1)}$$

   In this case, when the value of **ki** is high, the probability that it will be chosen to connect with the new node is higher.

2. If the new incoming node is a normal node, then the probability is adapted as follows:

$$\prod_{k_i} = \left(1-\frac{k_i}{k_{\max i}}\right)\frac{c_i}{\sum_{j\in local} c_j} \text{ ---------------------(2)}$$

   where $c_i$ is the number of cluster node *i*'s edges that should connect with the cluster node on both sides, and the greater the value of $c_i$ is, the higher the probability that it will be chosen to connect with the new node. Only through this approach we can adjust the number of cluster nodes that are linked to one cluster node (cluster head).

In [9], the authors consider the expenditure of energy in the process of linking nodes together. The disadvantage is that the energy in a cluster node will have exhausted in only several rounds. In fact, the energy consumption is relatively low; thus, the energy consumption is not considered in this stage, and only $k_{\max i}$ is considered to be the limit for a cluster node to connect to others randomly.

**Anti-Preferential Attachment:** Let us consider a parameter **z** called the deletion rate, which is defined as the rate of links removed divided by the rate of addition of links. It's observed that lesser the energy of the node, the more will be the probability of being deleted, let this probability be denoted as $\prod^{*}(k_i)$, now suppose that the sensor nodes which undergo a change in the configuration after time step **t** have sufficiently large scale we obtain the result for two type of nodes

For the outgoing cluster nodes, we have

$$\prod^{*}(k_i) \approx \frac{1}{m_0 + p^{*}t}.$$

For the outgoing normal nodes, we have

$$\prod^{*}(c_i) \approx \frac{1}{m_0 + (1-p)*t}.$$

So the total anti-preferential probability is

$$\prod^{*}(k_i) + \prod^{*}(c_i) = \prod^{*}$$

$$\prod^{*} \; = \frac{2\{(2*m_0/t)+1\}}{t*(\frac{m_0}{t}+p)*(\frac{m_0}{t}+1-p)}$$ (Factor 2 introduced due to mutual interaction)

--------------------(3)

The anti-preferential removal mechanism is more reasonable for deleting links that are anti-parallel with the preferential connection. It is consistent with the real wireless sensor networks environment. The wireless links that are not active may be removed from the network when the energy of the connecting nodes falls down to a certain level. The particular anti-preferential removal phenomenon is also reasonable for many real networks.

Now we adopt the form and usage mean field theory [10][11] to give a qualitative analysis for our energy-aware evolving model with link and node deletions. By the mean-field theory, let the following dynamical equation is as:

$$\frac{\delta k_i}{\delta t} = \frac{M}{m_0 + p*t} * \prod_{ki} \; - \; \frac{M*z}{m_0 + p*t}\left[\prod_{ki}^{*} + \sum_{i}\prod_{ki}^{*} * k_i^{-1} * \prod_{ki}^{*}\right]$$ --------------------- (4)

Using equations 1 , 2 , 3 we have,

$$\frac{\delta k_i}{\delta t} = \frac{M}{m_0 + p*t} * \left[ p*(1-\frac{k_i}{k_{\max i}})*\bar{k}_i^{-1} + (1-p)*(1-\frac{k_i}{k_{\max i}})*\bar{c}_i^{-1} \right] - \frac{2*M*z}{m_0 + p*t}\left[\frac{2*m_0 + t}{(m_0 + p*t)*(m_0 + t - p*t)}\right]$$

…………………………..(5)

Where

$$\frac{1}{\bar{k}_i} = \frac{k_i}{\sum_i k_i} \quad \text{and} \quad \frac{1}{\bar{c}_i} = \frac{c_i}{\sum_i c_i}$$

## 4. Analysis of the Dynamic Equation

**CASE:I**

If **z=0 , M=1** ie the new node selects node unless it reaches k. Moreover, the preferential attachment mechanism does not work. The rate of growth of $\mathbf{k_i}$ is as

$$\frac{\delta k_i}{\delta t} = \frac{1}{m_0 + p*t}$$

The denominator of the above expression is the number of cluster nodes at time **t.**

**CASE:II**

If $M=m_0+p*t$. This means that the local world is the whole network

$$\frac{\delta k}{\delta t} = p*(1-\frac{k_i}{k_{\max i}})*\overline{k^{-1}}_i + (1-p)*(1-\frac{k_i}{k_{\max i}})*\overline{c_i^{-1}} - \frac{2*M*z}{m_0+p*t}\left[\frac{2*m_0+t}{(m_0+p*t)*(m_0+t-p*t)}\right] \qquad \text{.........................(6)}$$

In a network, the degrees $k_i$ of most of the nodes are much smaller than their maximum $k_{\max i}$; thus, similarly, we obtain the following

$$1-\frac{k_i}{k_{\max i}} \approx 1 \qquad \textbf{...........................................................(7)}$$

i.e.

Putting the value of equation 7 in equation 6 we have

$$\frac{\delta k}{\delta t} = p*\frac{k_i}{\sum_i k_i} + (1-p)*\frac{c_i}{\sum_i c_i} - \frac{2*M*z}{m_0+p*t}\left[\frac{2*m_0+t}{(m_0+p*t)*(m_0+t-p*t)}\right] \text{...........(8)}$$

By definition

$$\overline{k} = \frac{\sum_i k_i}{k_i}$$

$$\overline{k} = total_degree_of_node/total_number_of_nodes$$

$$= \frac{m_0+p*t+N}{m_0+p*t} = \frac{m_0+p*t+m_0+t}{m_0+p*t}$$

Therefore $\overline{k} = \dfrac{2*m_0+t+p*t}{m_0+p*t}$ .............(9)

Similarly, we have for $c_i$

$$\overline{c} = \frac{\sum_j c_j}{c_j} = 2 \qquad \text{..................(10)}$$

(As the cluster node will have one such node attached to itself the status of that node is either another cluster head or normal head hence the count value of $\overline{c}$ is 2 )

Here the above equations (8) and (9) are used to find the values of $\overline{c}$ and $\overline{k}$ in fundamental terms then,

Finally the equation forms after substituting the values of $\sum_i c_i$ and $\sum_i k_i$ in equation 8 we have,

$$\frac{\delta k}{\delta t} = \frac{p*k_i}{2*m_0 + p*t + t} + (1-p)*\frac{c_i}{2*(m_0 + p*t)} - \frac{2*z*(2*m_0 + t)}{(m_0 + p*t)*(m_0 + t - p*t)}$$

At t-> infinity, $m_0$ ->0

$$\frac{\delta k}{\delta t} = \frac{p*k_i}{p*t + t} + (1-p)*\frac{c_i}{2*p*t} - \frac{2*z*t}{p*t*(t - p*t)}$$

Reducing the above equation by assuming **A=p/1+p B=1-p/p** , **C=1/p*(p-1**) and **$c_i$=2**

$$\frac{\delta k}{\delta t} = \frac{A*k_i}{t} + \frac{B}{t} - \frac{2*z*C}{t} \quad \ldots\ldots\ldots\ldots\ldots\ldots\ldots\ldots\ldots\ldots\ldots\ldots\ldots\ldots\ldots(10)$$

With Initial conditions are given as **$k_i(t_i)$=1** ,
By integration, we have the solution as

$$\frac{t}{t_i} = \left(\frac{k(t)*A + B - 2*z*C}{A + B - 2*z*C}\right)^{1/A}$$

Moreover, to find the degree distribution **P(k)** i.e., the probability that a node has **k** edges), we first calculate the cumulative probability **P[$k_i$(t)<k]** . Suppose that the node enters into the network in equal time intervals. We define the probability density **$t_i$** of as follows:

**P($t_i$)=1/$m_0$+t**

So **P[$k_i$(t)<k]** has the following form

$$(1 - \frac{t}{t + m_0})\left(\frac{k*A + B - 2*z*C}{A + B - 2*z*C}\right)^{1/A}$$

Hence, the degree distribution P(k) can be obtained:

$$\frac{\partial}{\partial t}\left[(1 - \frac{t}{t + m_0})\left(\frac{k*A + B - 2*z*C}{A + B - 2*z*C}\right)^{1/A}\right]$$

$$\approx (A + B - 2*C)^{1/A} * (k*A + B - 2*z*C)^{\frac{-1}{A}-1} \quad \ldots\ldots\ldots\ldots\ldots\ldots\ldots\ldots(11)$$

Equation 11 denotes the degree distribution function to understand the dynamics of clustering Putting back the values of A, B, C we have the distribution function as the degree distribution P(k) the probability that a node has k edges is

$$P(k) \approx \left(\frac{p}{p-1} + \frac{1-p}{p} - \frac{2*z}{p*(1-p)}\right)^{\frac{1+p}{p}} * \left(k*\frac{p}{1+p} + \frac{1-p}{p} - 2*\frac{z}{p*(1-p)}\right)^{\frac{-2*p-1}{p}}$$

………………(12)

Considering the equation (12) , an attempt to find the minimum value of **z** is done by considering the following equation below.

$$\frac{dP(k)}{dz} = 0 \text{ -------------- (13)}$$

By solving equation (13) we obtain the minimum value of **z** as

$$z = \frac{(1 + A + B - k)}{2 * C} \text{ ------------------(14)}$$

Now considering the anti-preferential factor (z) to be minimum, during the evolution of the network as the result the value of **z** will tend to zero.

By setting the value of **z** to zero we have the following relation

$$p = \frac{1}{2}\left[\left(\frac{k+3}{k-1}\right)^{0.5} - 1\right] \text{ ------------------- (15)}$$

## 5. Analysis of the Localized TC Protocols (A3, Simple tree Protocol, CDS-Rule K)

The simulation was carried using the Atarraya [16] Simulator where the total nodes collection considered was 100,200,300,500 respectively and they were tested for A3, Simple Tree, CDS-Rule K . The output was recorded for average **k** for the respective node collection considered for various above mentioned protocols.

The following diagram enumerates the investigation of average **k** value from the figure I

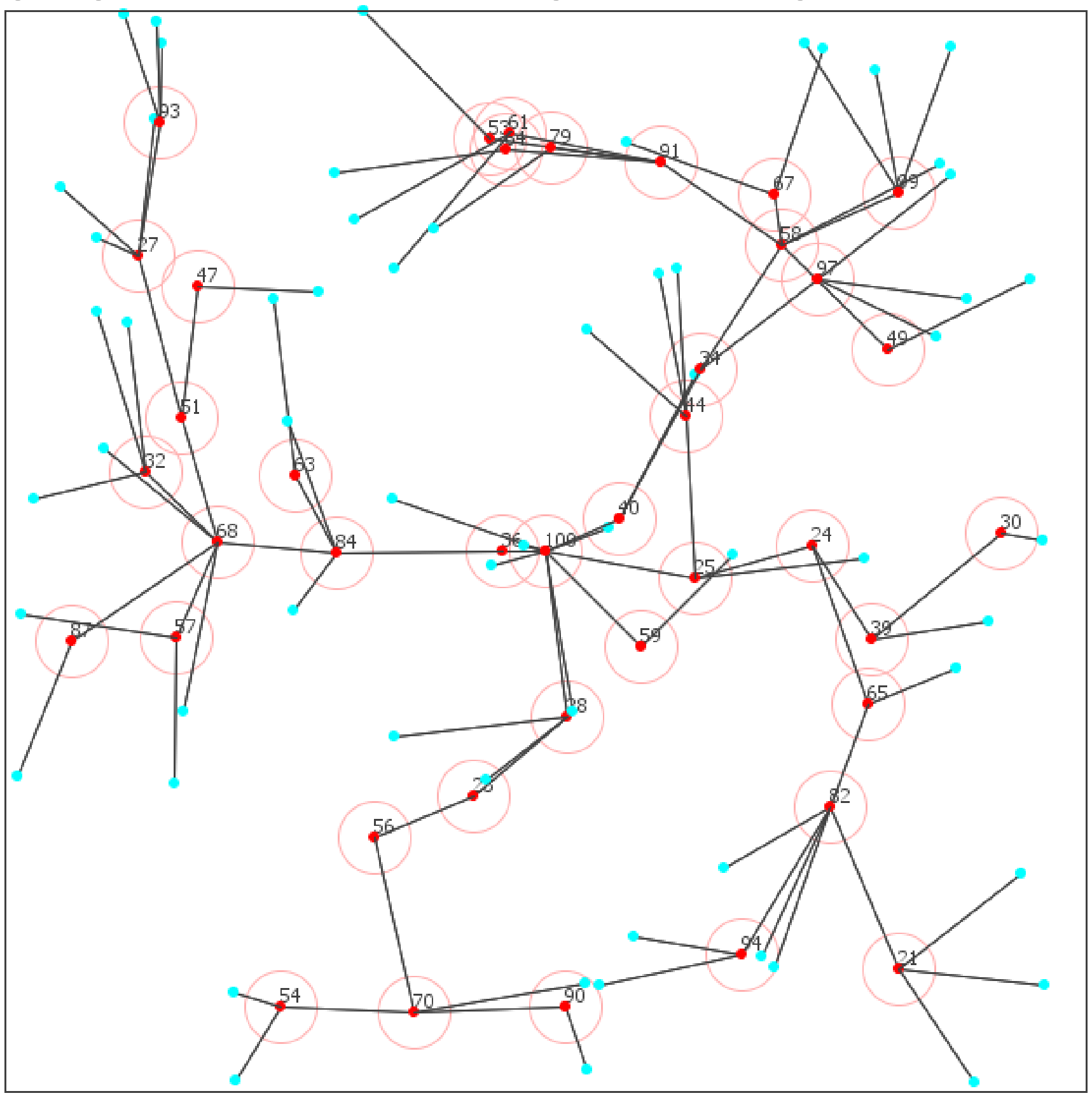

*Output of the simple tree protocol for total of 100 nodes*

**Figure-I**

**.**This figure shows the cluster heads in red circles with normal nodes attached. Taking the number of all the neighboring nodes of every cluster heads and dividing by the number of cluster heads we obtain the average value of **k.** By similar procedure we obtain the average value of **k** for other node collection viz. 200, 300,400,500.

Carrying out the simulation using Simple Tree Protocol for all collection then combining the theoretical (using equation 15) and experimental values we have the table as described below.

| N | k | n(T) | n(E) |
|---|---|---|---|
| 100 | 3 | 37 | 40 |
| 200 | 3 | 73 | 71 |
| 300 | 4 | 79 | 71 |
| 400 | 4 | 105 | 96 |
| 500 | 4 | 132 | 130 |

*N=collection, k=no. of neighbors, n(T)=Theoretical calculation of clusters, n(E)=Experimental number of clusters*

**Table-I**

Similarly we obtain for CDS-Rule k as

| N | k | n(T) | n(E) |
|---|---|---|---|
| 100 | 4 | 27 | 27 |
| 200 | 6 | 35 | 38 |
| 300 | 10 | 31 | 33 |
| 400 | 10 | 41 | 43 |
| 500 | 13 | 39 | 43 |

*N=collection, k=no. of neighbors, n(T)=Theoretical calculation of clusters, n(E)=Experimental number of clusters*

**Table-II**

Similarly we obtain for A3 protocol

| N | k | n(T) | n(E) |
|---|---|---|---|
| 100 | 4 | 27 | 34 |
| 200 | 6 | 35 | 40 |
| 300 | 8 | 39 | 49 |
| 400 | 10 | 41 | 48 |
| 500 | 12 | 42 | 46 |

*N=collection, k=no. of neighbors, n(T)=Theoretical calculation of clusters, n(E)=Experimental number of clusters*

**Table-III**

### A. Discussion on A3 and Simple Tree Protocol

Now we plot the Table I and III and we observe the following

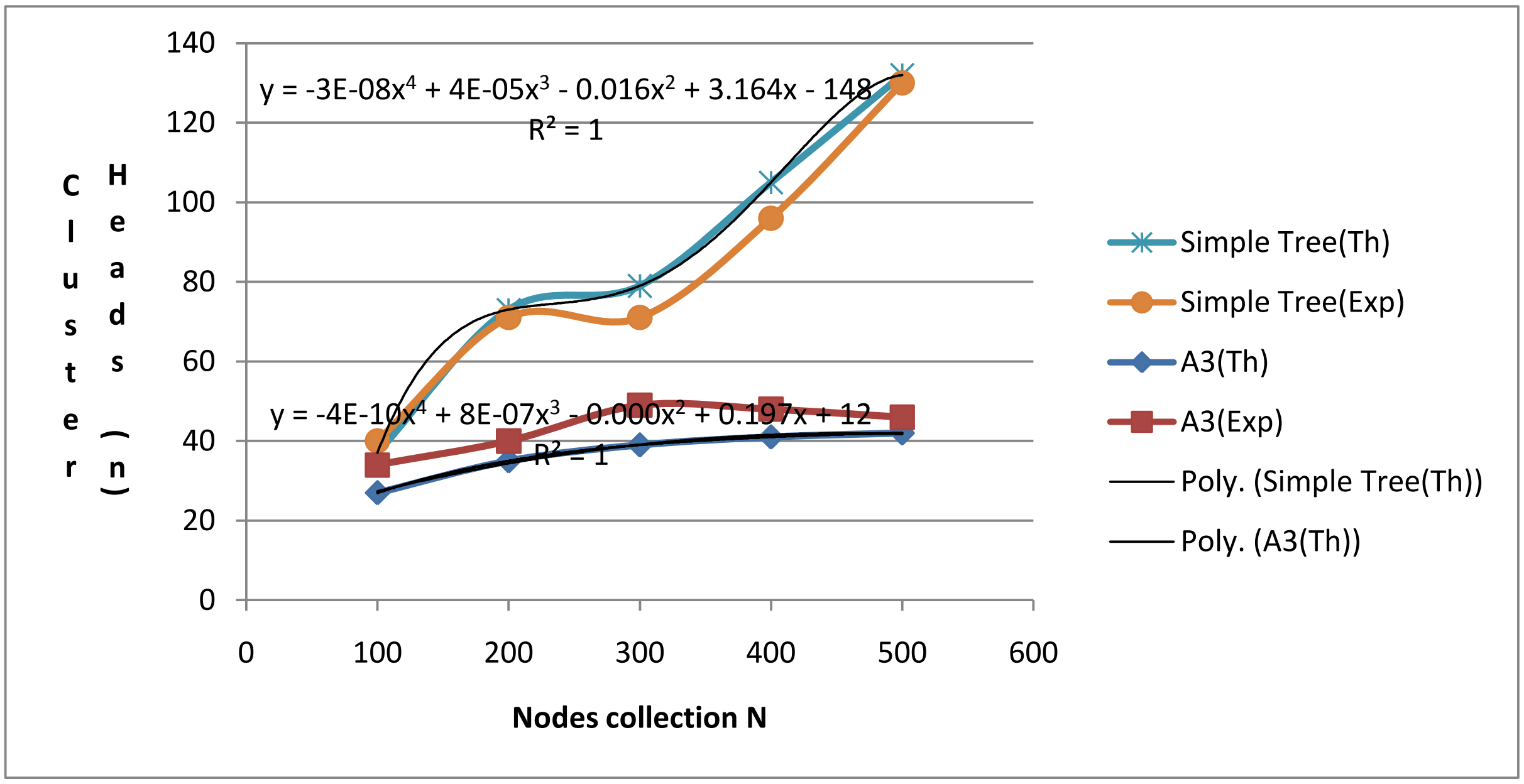

[*plot between the total nodes(N) (x) vs the cluster Nodes (n) (y)*]

**Figure-II**

1. The protocols (A3, Simple Tree) experimental as well as theoretical predicted curve follows the fourth degree equation (The trend line polynomial options of MS Excel have been used to study the degree of curve fit of the theoretical/experimental curve ).
2. There is flip between the theoretical and experimental curve in case of A3 and Simple Tree protocol which may be attributed due to the fact that A3 (approximate CDS) only preserves 1-connectivity whereas the Spanning Tree protocol has multiple connectivity, hence A3 protocol will require less energy to construct the tree as compared to span tree which confirms to the experimental results

**B. Discussion on CDS –Rule K**

Consider the equation (14) ie.

$z = \frac{(1 + A + B - k)}{2 * C}$ Where A, B, C, k, z have been defined previously. Table IV represents the calculated local minima for various values of k corresponding to the nodes collection.

| Nodes collection (N) | k | \|p\| | z |
|---|---|---|---|
| 100 | 4 | 0.5 | -7 |
| 200 | 6 | 0.5 | -9 |
| 300 | 8 | 0.5 | -11 |
| 400 | 10 | 0.5 | -13 |
| 500 | 13 | 0.5 | -16 |

**Table-IV**

Combining the results of Table II and Table IV we obtain the following graph as

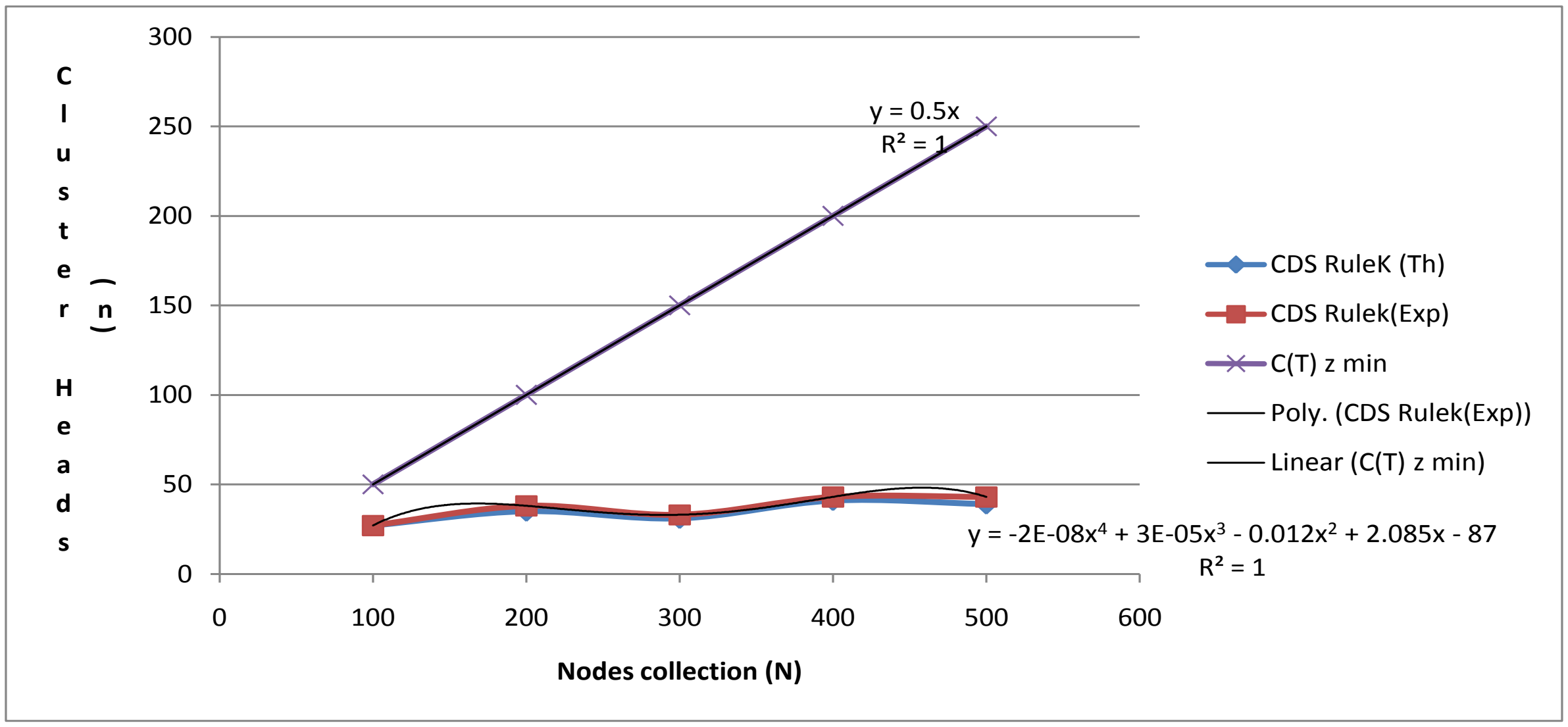

[*plot between the total nodes(N) (x) vs the cluster Nodes (n) (y)*]

**Figure-III**

In Table-IV the minimum value of z corresponds to a probability value 0.5 which indicates that the clusters have dissociated into one clusterhead and one normal node which ideally predicts the case when the energy of the nodes have drained out.

In Fig III, it can be seen that the mechanics of clustering in CDS- Rule K is in accordance to the assumptions considered while deducing the distribution function using mean field theory. Unlike A3 and Simple tree protocol, CDS-Rule K also exhibits a relation of degree 4 as shown. The linear graph as shown in the figure represents the situation when the value of **z** is min. This graph is linear due to constant value of **p** (0.5 here) where as other graphs were drawn on constant **z** (0 here).

Above experimental results lead us to the following explanation of the anti preferential factor **z.**

The anti preferential factor can have three distinct values

- **a.** When z=0 then the effective degree of the node is equal to the mathematical degree of a node. Here the number of cluster heads formed will be optimal.
- **b.** When z>0 then a node will have alternative connections available. Here the number of cluster heads will be higher.
- **c.** When z<0 then a node will have single connectivity. Here the number of cluster heads will be highest as the result will lead to dissociation.

## 6. Conclusions:

The assumptions work fairly well with the topology control (TC) protocols .The significance of using this approach will allow us to understand new approach in designing the TC protocols using the **z** factor other than the conventional big ohh (O) notation, especially the table IV which have far reaching consequences in understanding the lifetime of the network, if at all the average number of neighbor nodes per cluster head can be set through certain mechanism.

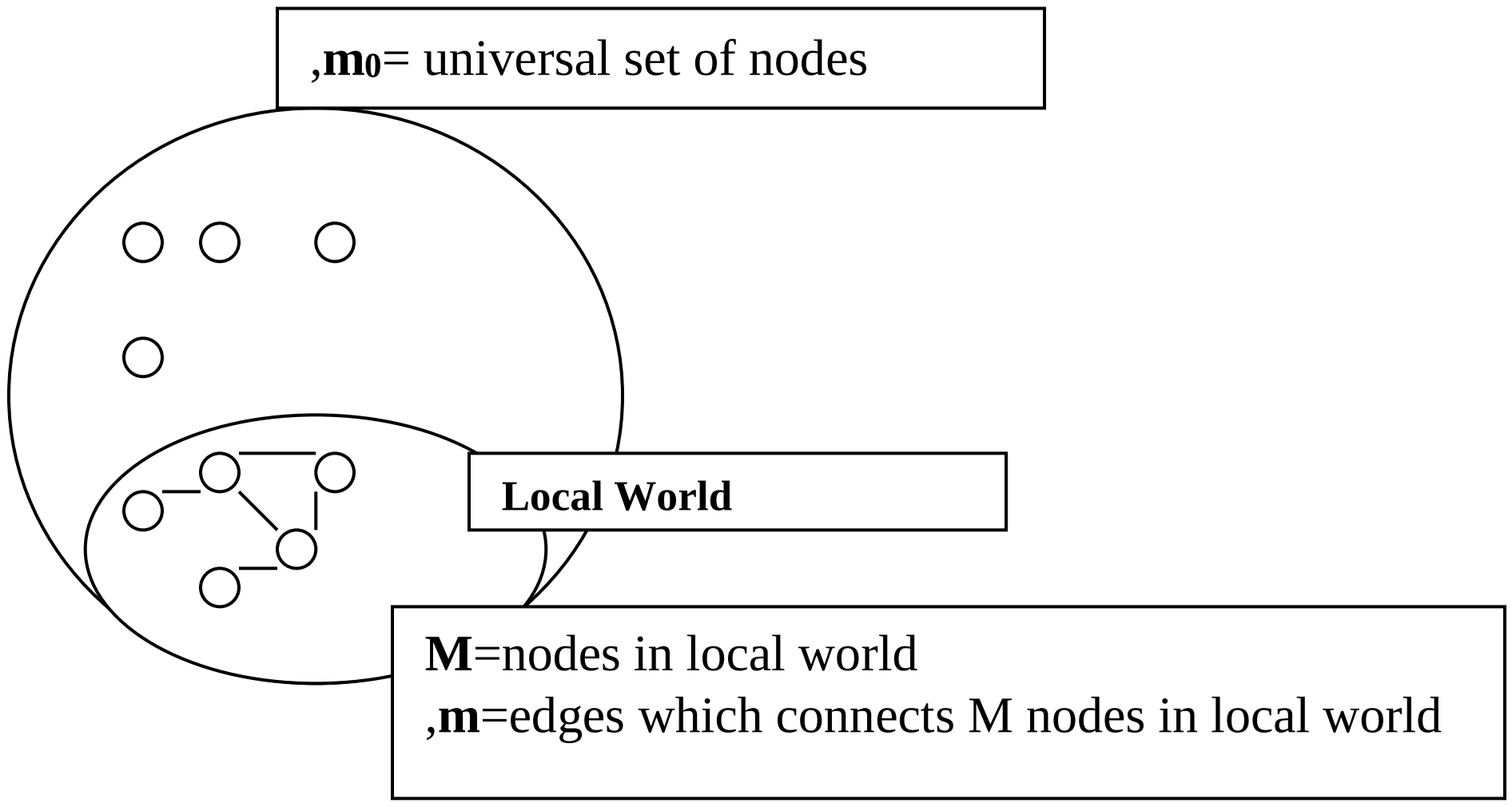